\documentclass[proof]{pasj01}

\begin{document} 
\Received{}%{yyyy/mm/dd}
\Accepted{}%{yyyy/mm/dd}
%\Published{yyyy/mm/dd}

\title{A Systematic Study of Evolved Supernova Remnants in the Large and Small Magellanic Clouds with Suzaku}

%%% begin:list of authors
% Do NOT capitalize all letters in "textsc".
\author{Yoko \textsc{takeuchi}\altaffilmark{1,2*},
Hiroya \textsc{yamaguchi}\altaffilmark{3,4},
and
Toru \textsc{tamagawa}\altaffilmark{2}
}
\email{takeuchi@crab.riken.jp}

\altaffiltext{1}{Department of Physics, Tokyo University of Science, 3-1 Kagurazaka, Shinjuku-ku, \\Tokyo 162-8601, Japan}
\altaffiltext{2}{RIKEN Nishina Center, 2-1 Hirosawa, Wako, Saitama, 351-0198, Japan}
\altaffiltext{3}{NASA Goddard Space Flight Center, Code 662, Greenbelt, MD 20771, USA}
\altaffiltext{4}{Department of Astronomy, University of Maryland, College Park, MD 20742, USA}

%%% end:list of authors

%% `\KeyWords{}' always has to be placed before `\maketitle'.
%%Do NOT move this preamble from here!

\maketitle

\begin{abstract}
Typing the origin (i.e., Type Ia or core-collapse) of supernova remnants (SNRs) is crucial to determine the rates
of supernova (SN) explosions in a galaxy, which is a key to understand its recent chemical evolution.
However, evolved SNRs in the so-called Sedov phase are dominated by the swept-up interstellar medium (ISM),
making it difficult to determine their ejecta composition and thus SN type. Here we present a systematic X-ray study
of nine evolved SNRs in the Magellanic Clouds, DEM\,L238, DEM\,L249, 0534--69.9, 0548--70.4, B0532--71.0,
B0532--67.5, 0103--72.6, 0049--73.6, and 0104--72.3, using archival data of the Suzaku satellite.
Although Suzaku does not spatially resolve the SN ejecta from the swept-up ISM due to the limited angular resolution,
its excellent energy resolution has enabled clear separation of emission lines in the soft X-ray band.
This leads to the finding that the `spatially-integrated' spectra of the evolved ($\sim 10^4$\,yr) SNRs
are still significantly contributed by emission from the ejecta at the energies around 1 keV.
The Fe/Ne mass ratios, determined mainly from the well-resolved Fe L-shell and Ne K-shell lines,
clearly divide the observed SNRs into the Type Ia and core-collapse groups, confirming some previous
typing made by Chandra observations that had utilized its extremely high angular resolution.
This demonstrates that spatially-integrated X-ray spectra of old SNRs can also be used to discriminate their progenitor type,
which would be helpful for future systematic studies of extragalactic SNRs with ASTRO-H and beyond.
\end{abstract}

\begin{table*}[t]
\begin{center}
\tbl{List of the Magellanic SNRs we analyze in this work.}{
  \label{tab:fit}
\begin{tabular}{lcclccccc}
\hline 
\hline 
Name	&  Suzaku~Obs. ID        	& Exposure	&Date of Observation& \multicolumn{2}{c}{Position (J2000)}     	& Radius		& Age	& Ref.\footnotemark[$*$] \\
			&			&(ks)			&				 &R.A. 	& Decl.				      	&  (arcsec / pc)	& (kyr)	& \\
\hline 
\multicolumn{7}{l}{LMC SNRs: } \\ 
~~~~DEM L238\footnotemark[$\dagger$]	& 505063010	& 200	&~~2010 Apr 22	&$05^{\rm h}34^{\rm m}24^{\rm s}$ 	&$-70^{\rm d}33^{\rm m}12^{\rm s}$	& 80/ 20				& 13.5 	&  1\\	
~~~~DEM L249\footnotemark[$\dagger$]	& 505063010	& 200	&~~2010 Apr 22	&$05^{\rm h}36^{\rm m}13^{\rm s}$ 	&$-70^{\rm d}38^{\rm m}10^{\rm s}$	& $\sim$80/ $\sim$20	& 10$-$15	&  1\\	
~~~~0534--69.9 		& 505064010	& 109  	&~~2010 Mar 31	&$05^{\rm h}34^{\rm m}01^{\rm s}$ 	&$-69^{\rm d}54^{\rm m}22^{\rm s}$	& 57 / 13.7 	&  10.1		&  2\\
~~~~0548--70.4 		& 505065010	& 104  	&~~2010 July 1		&$05^{\rm h}74^{\rm m}52^{\rm s}$ 	&$-70^{\rm d}25^{\rm m}01^{\rm s}$	& 51 / 12.1 	&  7.1		&  2\\
~~~~B0532--71.0		& 803038010  	& 49		&~~2008 Apr 28	&$05^{\rm h}31^{\rm m}59^{\rm s}$ 	&$-71^{\rm d}00^{\rm m}03^{\rm s}$	& 87/ 21		& 23$-$27		& 3\\
~~~~B0532--67.5\footnotemark[$\ddagger$]	& 806007010	& 82		&~~2011 Nov 6		&$05^{\rm h}32^{\rm m}23^{\rm s}$ 	&$-67^{\rm d}31^{\rm m}16^{\rm s}$	& 48/11.6		& ---			& --- \\	
\hline 
\multicolumn{7}{l}{SMC SNRs: } \\
~~~~0103--72.6 		& 501077010	& 49		&~~2006 Apr 23	&$01^{\rm h}05^{\rm m}07^{\rm s}$ 	&$-72^{\rm d}23^{\rm m}10^{\rm s}$	& 85 / 24.7	&18 	&  4\\
~~~~0049--73.6		& 503094010 	& 120	&~~2008 June 12	&$00^{\rm h}51^{\rm m}09^{\rm s}$ 	&$-73^{\rm d}21^{\rm m}54^{\rm s}$	& 72 / 20.5	&14	&  5\\
~~~~0104--72.3 		& 803002010 	&107		&~~2008 May 15	&$01^{\rm h}06^{\rm m}19^{\rm s}$ 	&$-72^{\rm d}05^{\rm m}41^{\rm s}$	& 55 / 16		&17.6& 6\\
\hline 	
\end{tabular}}
\label{tab:SNR_list}
\end{center}
\begin{tabnote}
\footnotemark[$*$] Representative references --- (1)~\citet{Borkowski+06}: (2)~\citet{Hendrick+03}: (3)~\citet{Williams+05}: (4)~\citet{Park+03}: (5)~\citet{Hendrick+05}:  (6)~\citet{Lopez+14}\\
\footnotemark[$\dagger$]	These two SNRs were observed in the same FoV. \\
\footnotemark[$\ddagger$]	The age of B0532--67.5 is previously unreported. \\
\end{tabnote}
\end{table*}

\section{Introduction}
\label{sec:intro}

A supernova (SN) explosion provides heavy elements synthesized during the stellar evolution 
or the explosion itself to the interstellar space efficiently. There are two distinct types of SNe: 
thermonuclear (Type Ia) explosions of a white dwarf(s) and core-collapse (CC) explosions of a massive star. 
Since the major nucleosynthesis products of both SN types are largely different from each other, it is crucial 
to determine their explosion rates in an individual galaxy to reveal its recent chemical enrichment history. 
X-ray observations of young supernova remnants (SNRs) allow us to distinguish their parent SN types, 
as their X-ray spectrum is dominated by emission from the ejecta (e.g., \cite{Vink+12}). 
The SN rate in the Large Magellanic Cloud (LMC), the nearest extra-galaxy to us and hence 
the best object for a comprehensive study of a galactic chemical evolution, was estimated 
using ASCA data of young ($t \lesssim 1,500$\,yr) LMC SNRs \citep{Hughes+95}.

On the other hand, typing the origin of an evolved ($t \gtrsim 5,000$\,yr) SNR is often very challenging, because 
the X-ray spectrum integrated from the entire SNR is dominated by the swept-up interstellar medium (ISM). 
In fact, previous ASCA and XMM-Newton observations of Magellanic Sedov-phase SNRs were utilized for 
studying the ISM abundances (e.g., \cite{Hughes+98}; \cite{van-der-Heyden+04}). 
%Unlike young SNRs, however, the reverse shock in evolved SNRs has reached at
%the remnant center, so that the entire SN ejecta are hot enough to emit thermal X-rays.
A breakthrough was achieved by the extremely high angular resolution ($\sim 0.\!''5$) of the Chandra 
X-ray Observatory. Allowing us to spatially resolve the ejecta core from the swept-up ISM shell, Chandra has 
successfully distinguished the progenitor types of a number of evolved SNRs in the Magellanic Clouds 
(e.g., DEM\,L71: \cite{Hughes+03}; N49B: \cite{Park+03_N49B}; N132D: \cite{Borkowski+07}).

However, opportunities for the spatially-resolved analysis are limited, since it requires both an expensive telescope 
with an adequate angular resolution and a target with a reasonable angular size. Although the forthcoming X-ray 
mission ASTRO-H \citep{Takahashi+14} will be dramatically changing our view of SNRs 
with its excellent spectral resolution ($\Delta E \sim 5$\,eV at 0.5--10\,keV), 
%enable detailed spectral studies on both Galactic and Magellanic SNRs. 
its half power diameter of $1.\!'3$ is larger than the typical angular size of evolved SNRs in 
the Magellanic Clouds ($\sim1'$). Moreover, SNRs in nearby spiral galaxies (e.g., M31) cannot be resolved 
into ejecta and ISM components even with the Chandra's angular resolution. It is, therefore, important to 
identify a simple (but reliable) diagnostic to distinguish SN types from integrated X-ray spectra alone. 
It would be worth noting that recent Suzaku observations of the middle-aged ($t \sim 5,000$\,yr) LMC SNRs 
DEM\,L71 and N49B robustly confirmed their abundance enhancement (Fe and Mg, respectively), 
which was originally revealed by the preceding Chandra observations, even from the spectra of 
the entire remnants \citep{Uchida+15}. This was enabled due to the high sensitivity 
(to emission lines) of the X-ray Imaging Spectrometer (XIS: \cite{Koyama+07}) on board Suzaku.

In this paper, we focus on even more evolved ($t \gtrsim 10,000$\,yr) SNRs in the LMC and 
the Small Magellanic Clouds (SMC), DEM\,L238, DEM\,L249, 0534--69.9, 0548--70.4, B0532--71.0, 
B0532--67.5, 0103--72.6, 0049--73.6, and 0104--72.3 (table\,\ref{tab:SNR_list}), observed by the Suzaku XIS. 
The progenitor type of most of these objects has been proposed by previous Chandra studies 
based on their spatially-resolved spectra (see \S4.1).
The aim of our present work is not to investigate the detailed nature of each individual object
nor to understand the detailed evolutionary characteristics of Sedov-phase SNRs, but to demonstrate 
that spatially-integrated X-ray spectra of the evolved SNRs can still be used to distinguish their progenitor type, 
based on an independent analysis of the high-quality Suzaku data. This not only strengthens (or competes with) 
the previous typing made by Chandra but also identifies a key spectral feature that immediately 
distinguishes an SN type, which should be useful for the future spectroscopy of extra galaxies. 
To achieve this goal, we systematically analyze the nine evolved SNRs with uniform data reduction and 
spectral models. Such an approach (i.e., uniform analysis) is known to be a reliable way for 
the progenitor type discrimination (e.g., \cite{Lopez+11}; \cite{Yamaguchi+14_FeKa}).

In \S2, we describe the observations and data reduction.
The results of the spectral analysis are presented in $\S$3, and discussed in \S4. 
Finally, we conclude this work in \S5.  
We assume 50\,kpc and 60\,kpc for the distances to the LMC and SMC, respectively \citep{Westerlund+90}.

%%%+++++++++++++++++++++++++++++++++++++
\section{Observations and Data Reduction}
\label{sec:obs}
We analyzed archival data of the nine SNRs taken by the Suzaku/XIS, 
of which observation details are summarized in table\,\ref{tab:SNR_list}. 
The XIS consists of four X-ray charge-coupled devices (CCDs). Three of them (XIS0, XIS2, and XIS3) are 
front-illuminated (FI) and the other (XIS1) is back-illuminated (BI).
The former has a better energy response and a lower background level, while the latter has superior 
sensitivity in the soft X-ray band. 
Combined with X-Ray Telescopes (XRTs; \cite{Serlemitsos+07}), the field of view (FoV) of the XISs, 
which are identical among the four chips, covers an $\sim$$18' \times 18'$ region with a half-power diameter 
(HPD) of $\sim$$2'$. 
We used all the CCD data for the analysis, but the XIS2 was out of operation during the observations 
of other than 0103--72.6, possibly caused by the impact of a micrometeorite. 

For data reduction, we used the HEAsoft version 16.6 software package. 
We reprocessed the data using the {\tt aepipeline} task with the latest calibration data released in May 2015.
The good time interval screening was performed in accordance with the standard criteria, 
obtaining the effective exposures given in table\,\ref{tab:SNR_list}.

For each SNR, we extracted XIS spectra from a circular region centered at the source with a diameter of 
$\sim$$4'$. A background spectrum was extracted from an annular region surrounding the source 
with inner and outer diameters of $12'$ and $16'$, respectively. 
We attempted several different background regions, and confirmed no significant change in the measured 
spectral parameters. We created XIS redistribution matrix files (RMFs) and ancillary response files (ARFs) 
using the {\tt xisrmfgen} and {\tt xissimarfgen} tasks. 
The following spectral analysis was performed with the XSPEC software version 12.8.2 \citep{Arnaud+96}.

%%%-------------------------------
\begin{figure*}[t]
  \begin{center}
        \includegraphics[width=18cm]{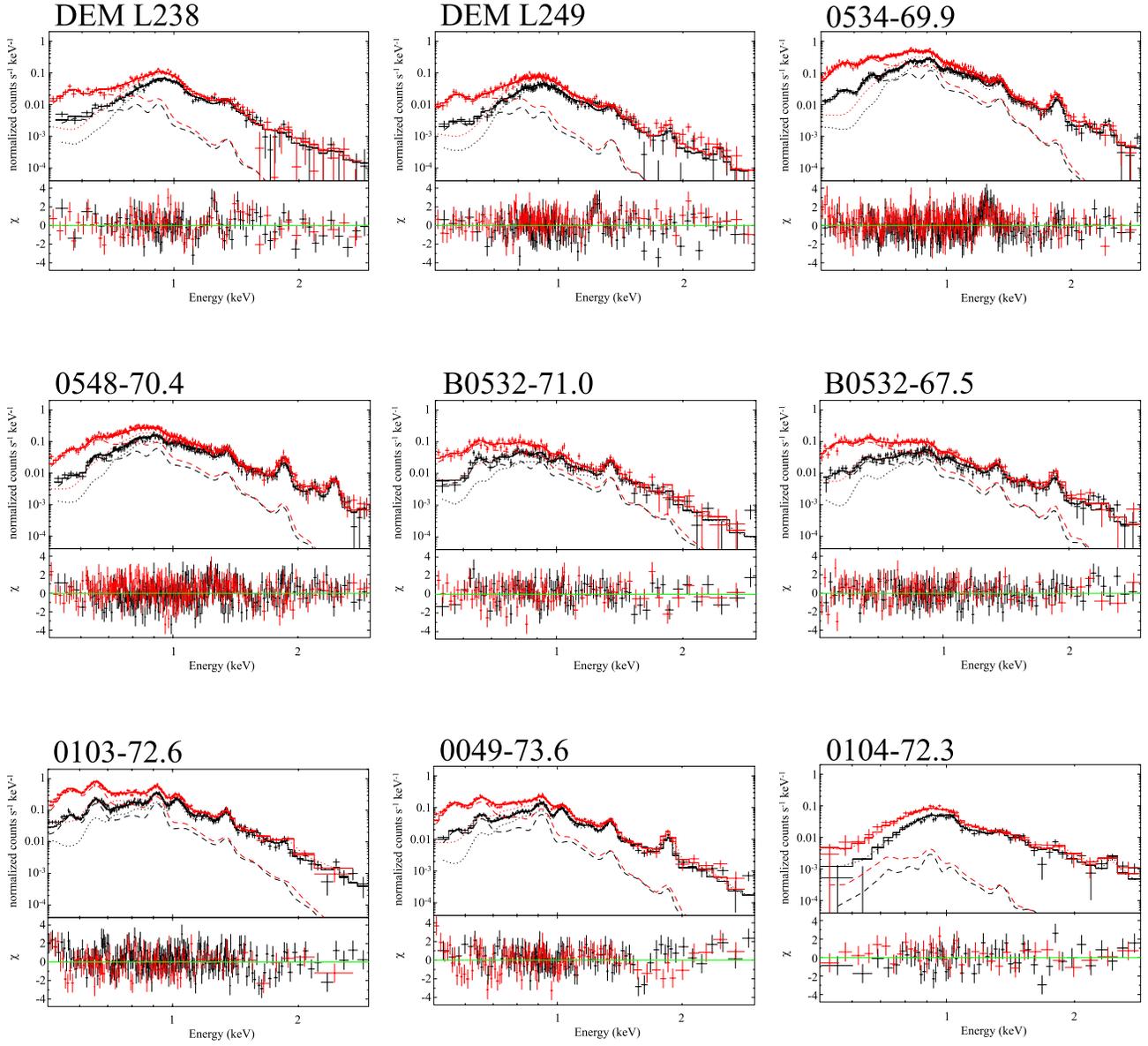} 
  \end{center}
\caption{
Suzaku~XIS spectra of the Magellanic SNRs fitted with the best-fit models given in table\,\ref{tab:fit}. 
The black and red data points represent the XIS FI and BI CCDs, respectively. 
The best-fit models are indiscated by the dashed and dotted lines.
}
\label{fig:spec}
\end{figure*}

%%%+++++++++++++++++++++++++++++++++++++
\section{Analysis and Results}
\label{sec:result}
Figure\,\ref{fig:spec} shows the XIS spectra of all the SNRs. The K-shell emission of 
O, Ne, Mg, and Si, and/or the Fe L-shell emission are clearly separated in most spectra. 
%A broad feature of Fe L-shell emission is enhanced in some of them (e.g., DEM\,L238), 
%whereas K-shell emission of O, Ne, and Mg is clearly resolved in others (e.g., 0103--72.6).
Since none of the SNRs shows a significant signal above $\sim$3\,keV, we focus on the 0.5--3.0\,keV data 
for the following spectral analysis. Here we perform an unbiased spectral modeling although the previous 
Chandra observations had revealed a degree of chemical inhomogeneity in some  SNRs. 
We first fit the spectra with a single-component variable-abundance non-equilibrium ionization (VNEI) 
plasma model based on the {\it AtomDB}\footnote{http://www.atomdb.org/index.php} version 3.0.2. 
The free parameters were the electron temperature ($kT_{\rm e}$), ionization parameter ($n_{\rm e}t$, 
where $n_{\rm e}$ and $t$ are the electron density and elapsed time since the gas was shock heated), 
emission measure (EM), and the abundances of O, Ne, Mg, Si, S, and Fe.
For several SNRs, however, we fixed the O, Si, and/or S abundances to the mean LMC/SMC ISM values of \citet{Russell+92} or \citet{Lewis+03}, because otherwise the fit did not converge. 
The abundances of the other elements were also fixed to the mean ISM values, but the Ni was tied to Fe. 
The interstellar extinction in the Galaxy and Magellanic Clouds was separately considered. 
The Galactic absorption column density with the solar abundances was fixed 
at $N_{\rm H}^{\rm Gal}$ = $6 \times 10^{20}$\,cm$^{-2}$, based on the Galactic H{\footnotesize I} 
observations of \citet{Dickey+90}. 
The other component ($N_{\rm H}^{\rm MC}$) was a free parameter, with the assumption of 
the LMC/SMC metal abundances (using the {\it phabs} model in XSPEC).

This model and assumptions yielded reasonable fits with $\chi ^2_{\nu}$ $\sim$ 1.2--2.0, 
but introducing a two-temperature NEI model resulted in a significantly better fits for all the SNRs 
other than 0104--72.3. 
We allowed $kT_{\rm e}$ and $n_{\rm e}t$ of the two components to vary independently, 
but the elemental abundances were linked between them. 
Typical electron temperatures of the high-$kT_e$ and low-$kT_e$ components were 
0.5--1.0\,keV and 0.2--0.3\,keV, respectively. The ionization parameter of the low-$kT_{\rm e}$ 
component was derived to be very high ($\gtrsim 10^{12}$\,cm$^{-3}$\,s) in most SNRs, 
suggesting that this component is almost in the collisional ionization equilibrium (CIE). 
In this case, we fixed $n_{\rm e}t$  to $1\times 10^{13}$\,cm$^{-3}$\,s, where the plasma 
is considered to be in the full CIE state. The best-fit parameters we obtained are given in table\,\ref{tab:fit}. 
Only in 0104--72.3, the second component with a free electron temperature did not improve the fit at all. 
Nevertheless, we attempted modeling with the two components by fixing the low-$kT_{\rm e}$ value to 0.3\,keV, 
to determine the upper limit of the low-$kT_{\rm e}$ emission measure.
We give in table 2 the results from both 1-$kT_{\rm e}$ and 2-$kT_{\rm e}$ modelings, showing no difference in
the derived elemental abundances. Similarly, we confirmed that the 1-component fit did not
alter the abundance pattern for the other SNRs when this model gave reasonable results
with $\chi^2_{\nu} \lesssim 1.5$.
We also fit the data of all the SNRs assuming the LMC/SMC abundances for the low-$kT_{\rm e}$ component, 
but the abundances of the other component did not change significantly from the values listed in table\,\ref{tab:fit}.

The best-fit models and residuals are shown in figure\,\ref{fig:spec}. 
In the first four spectra (DEM\,L238, DEM\,L249, 0534--69.9, and 0548--70.4), 
apparent disagreement between the data and model is seen at the energies around 1.2\,keV. 
Given that the large residuals are found only in the Fe-rich remnants (see table\,\ref{tab:fit}), this feature is 
likely to be associated with Fe L-shell emission. In fact, similar residuals were reported in a number of previous 
works (e.g., \cite{Brickhouse+00}; \cite{Uchida+15}), and interpreted as Fe L-shell 
transitions from high quantum numbers ($n > 5$) which are missing from the present plasma code. 
If we add a Gaussian to compensate these missing lines, the $\chi ^2_{\nu}$ values are significantly reduced, 
but no change is confirmed in the other best-fit parameters.

%%%--------------------------------------
\begin{table*}[t]
\begin{center}
\caption{The best-fit spectral parameters}
\label{tab:fit}
\tiny
\begin{tabular}{lcccccccccc}
\hline \hline
Parameters                &DEM L238              &DEM L249                &0534--69.9 &0548--70.4             & B0532--71.0 &B0532--67.5            &0103--72.6 &0049--73.6   &    0104--72.3  (1-$kT_{\rm e}$) &    0104--72.3  (2-$kT_{\rm e}$)    \\
\hline
~~~~$N_{\rm H}^{\rm MC}$\footnotemark[$*$]	&	0.0$_{-0.0}^{+6.3}$	&	4.3$_{-4.3}^{+13.8}$	&	34.2$_{-3.6}^{+11.8}$	&	49$_{-11}^{+12}$	&	0.16$_{-0.16}^{+9.07}$	&	8.2$_{-8.2}^{+22.5}$	&	0.0$_{-0.0}^{+1.2}$	&	0.0$_{-0.0}^{+1.3}$	&	0$_{-0}^{+13}$	&	0.45$_{-0.45}^{+12.83}$	\\
\multicolumn{11}{l}{(High temperature component)} \\
~~~~$kT_{\rm e}$   	&	1.05$_{-0.09}^{+0.10}$	&	0.86$_{-0.04}^{+0.14}$	&	0.68$_{-0.03}^{+0.03}$	&	0.68$_{-0.04}^{+0.04}$	&	0.63$_{-0.12}^{+0.15}$	&	0.73$_{-0.18}^{+0.25}$	&	0.52$_{-0.1}^{+0.04}$	&	0.63$_{-0.03}^{+0.03}$	&	0.87$_{-0.02}^{+0.02}$	&	0.88$_{-0.03}^{+0.05}$	\\
~~~~$n_{\rm e}t$\footnotemark[$\dagger$]	&	1.16$_{-0.46}^{+0.72}$	&	2.06$_{-1.12}^{+1.58}$	&	1.47$_{-0.32}^{+0.48}$	&	1.53$_{-0.39}^{+0.99}$	&	2.44$_{-1.05}^{+9.21}$	&	1.09$_{-0.51}^{+3.37}$	&	3.34$_{-1.12}^{+4.84}$	&	1.81$_{-0.33}^{+0.51}$	&	100 (fixed)	&	100 (fixed)	\\
~~~~EM\footnotemark[$\ddagger$]	&	2.77$_{-0.90}^{+1.14}$	&	2.35$_{-0.90}^{+0.99}$	&	20.4$_{-2.9}^{+3.9}$	&	26.0$_{-3.7}^{+5.0}$	&	13.4$_{-4.6}^{+10.2}$	&	10.2$_{-3.9}^{+7.2}$	&	54$_{-11}^{+35}$	&	11.3$_{-2.3}^{+2.6}$	&	9.4$_{-2.2}^{+2.1}$	&	9.2$_{-2.1}^{+2.2}$	\\
\multicolumn{11}{l}{(Low temperature component)} \\
~~~~$kT_{\rm e}$   	&	0.172$_{-0.005}^{+0.005}$	&	0.173$_{-0.012}^{+0.018}$	&	0.169$_{-0.008}^{+0.003}$	&	0.181$_{-0.006}^{+0.007}$	&	0.259$_{-0.034}^{+0.035}$	&	0.277$_{-0.089}^{+0.153}$	&	0.202$_{-0.009}^{+0.011}$	&	0.186$_{-0.004}^{+0.007}$	&	---	&	0.3 (fixed)	\\
~~~~$n_{\rm e}t$\footnotemark[$\dagger$]	&	100 (fixed)	&	10.24$\pm$9.3	&	100 (fixed)	&	100 (fixed)	&	100 (fixed)	&	0.89$_{-0.84}^{+3.3}$	&	100 (fixed)	&	100 (fixed)	&	---	&	100 (fixed)	\\
~~~~EM\footnotemark[$\ddagger$] &	31$_{-16}^{+27}$	&	26$_{-15}^{+48}$	&	670$_{-250}^{+590}$	&	310$_{-130}^{+200}$	&	41$_{-15}^{+21}$	&	33$_{-29}^{+157}$		&153$_{-23}^{+30}$	&	52$_{-11}^{+14}$	&	---	&	0.66$_{-0.66}^{+3.94}$	\\
\multicolumn{11}{l}{(Abundances)} \\
~~~~O	&	0.22$_{-0.12}^{+0.24}$	&	0.25$_{-0.13}^{+0.38}$	&	0.29$_{-0.08}^{+0.06}$	&	0.31$_{-0.04}^{+0.13}$	&	0.263 (fixed)	&	0.263 (fixed)	&	0.36$_{-0.06}^{+0.07}$	&	0.51$_{-0.11}^{+0.15}$	&	0.126 (fixed)	&	0.126 (fixed)	\\
~~~~Ne	&	0.57$_{-0.32}^{+0.38}$	&	0.47$_{-0.36}^{+0.41}$	&	0.41$_{-0.09}^{+0.11}$	&	0.38$_{-0.07}^{+0.1}$	&	0.25$_{-0.06}^{+0.12}$	&	0.27$_{-0.04}^{+0.09}$	&	0.62$_{-0.06}^{+0.13}$	&	1.09$_{-0.19}^{+0.27}$	&	1.5$_{-0.59}^{+0.81}$	&	1.47$_{-0.77}^{+0.84}$	\\
~~~~Mg	&	0.60$_{-0.26}^{+0.50}$	&	0.96$_{-0.45}^{+0.99}$	&	0.52$_{-0.09}^{+0.13}$	&	0.45$_{-0.08}^{+0.14}$	&	0.61$_{-0.15}^{+0.12}$	&	0.27$_{-0.07}^{+0.08}$	&	0.37$_{-0.06}^{+0.09}$	&	0.71$_{-0.14}^{+0.18}$	&	0.57$_{-0.26}^{+0.39}$	&	0.57$_{-0.27}^{+0.4}$	\\
~~~~Si	&	0.18$_{-0.18}^{+0.3}$	&	0.74$_{-0.25}^{+0.81}$	&	0.76$_{-0.15}^{+0.19}$	&	0.97$_{-0.15}^{+0.25}$	&	0.309 (fixed)	&	0.66$_{-0.16}^{+0.25}$	&	0.21$_{-0.07}^{+0.06}$	&	1.25$_{-0.24}^{+0.3}$	&	0.26$_{-0.17}^{+0.23}$	&	0.26$_{-0.18}^{+0.23}$	\\
~~~~S	&	0.309 (fixed)	&	1.12$_{-0.91}^{+0.94}$	&	0.70$_{-0.31}^{+0.33}$	&	1.75$_{-0.34}^{+0.49}$	&	0.309 (fixed)	&	0.59$_{-0.45}^{+0.72}$	&	0.24 (fixed)	&	0.24 (fixed)	&	1.24$_{-0.53}^{+0.72}$	&	1.24$_{-0.53}^{+0.72}$	\\
~~~~Fe	&	1.62$_{-0.48}^{+0.81}$	&	1.96$_{-0.65}^{+1.13}$	&	1.12$_{-0.18}^{+0.23}$	&	0.60$_{-0.11}^{+0.17}$	&	0.17$_{-0.05}^{+0.06}$	&	0.16$_{-0.04}^{+0.05}$	&	0.10$_{-0.04}^{+0.01}$	&	0.32$_{-0.07}^{+0.09}$	&	0.36$_{-0.07}^{+0.12}$	&	0.36$_{-0.08}^{+0.13}$	\\
~~~~Flux\footnotemark[$\S$]  &5.56 	& 5.00  & 27.5 	&14.5	&	8.75 	&	7.73  &	27.6	&	12.9	&	3.51&	3.53	\\
~~~~$L_{0.5-3 \rm keV}$\footnotemark[$\|$] & 1.66	&	1.50	&	8.23	&	4.34	&	2.62	&	2.31	&	11.9	&	5.56	&	1.51	&	1.52	\\
~~~~$\chi^2_{\nu} (d.o.f)$	&	1.41 (168)	&	1.25 (261)	&	1.26 (478)	&	1.22 (465)	&	1.16 (170)	&	1.00 (240)	&	1.31 (323)	&	1.30 (208)	&	1.19(88)	&	1.20 (87)	\\
\hline
\end{tabular}
\end{center}
\begin{tabnote}
The uncertainties are in the 90\% confidence range.\\
\footnotemark[$*$] The unit is 10$^{20}$ cm$^{-2}$.\\
\footnotemark[$\dagger$] The unit is 10$^{11}$ cm$^{-3}$\,s.\\
\footnotemark[$\ddagger$] The unit is 10$^{57}$ cm$^{-3}$.\\
\footnotemark[$\S$] The unit is $10^{-13}$ ergs s$^{-1}$ cm$^{-2}$.\\
\footnotemark[$\|$] The unit is $10^{35}$ ergs s$^{-1}$.\\
The Galactic absorption ($N_{\rm H}^{\rm Gal}$) was fixed to 6$\times10^{20}~$cm$^{-2}$.
\end{tabnote}
\end{table*}

%%%+++++++++++++++++++++++++++++++++++
\section{Discussion}
\label{sec:result}

We have performed systematic studies of the nine evolved SNRs in the Magellanic Clouds using 
the spatially-integrated spectra of the Suzaku XIS with uniform data reduction and spectral modeling. 
Despite the simpleness of our spectral modeling, we have been able to obtain the reasonable fits and 
the constraints on elemental abundances of the individual heavy elements. 
We have found the significant variation in the abundance pattern among the objects. 
In this section, we first describe a brief summary of the previous works to compare with our results, 
and then identify the best spectral feature to discriminate the progenitor type of the evolved SNRs 
(without spatially-resolved analysis). Finally, we discuss future prospects for ASTRO-H and beyond.

\subsection{Summary of the Preceding Works and Comparison with Our Results}
\label{sec:sum}

DEM L238 \& DEM L249:
Both SNRs were studied in detail with Chandra and XMM-Newton by \citet{Borkowski+06}. 
Using the Chandra high-resolution data, they revealed that prominent Fe L-shell emission 
dominates the spectra from the SNR center, which suggests the SN Ia origin for the both.
Our analysis also confirms the enhanced Fe abundances, whereas the abundances of 
the lighter elements (i.e., O, Ne, and Mg) are comparable to the mean LMC values. 
On the other hand, the ionization parameters we obtained (for the high-$kT_e$ component) 
are significantly lower than the previous measurements by \citet{Borkowski+06} ($\sim 10^{12}$\,cm$^{-3}$\,s).

0534--69.9 \& 0548--70.4:
The first detailed studies of these SNRs were made by Chandra observations \citep{Hendrick+03}.
Similarly to DEM L238 and DEM L249, the SNR center spectra indicate the strong Fe L emission. 
Balmer-dominated optical spectra of 0548--70.4 also suggest its SN Ia origin \citep{Smith+91}. 
However, the O/Fe number ratio observed in this SNR is comparable to or higher than the mean 
LMC value \citep{Hendrick+03}, which is unusual for SN Ia ejecta composition. 
Our results confirms the strong Fe L emission, and reveal that the O/Fe ratio is significantly lower 
than the previous measurement, now reasonable as an evolved Type Ia SNR.

B0532--71.0: 
This SNR is associated with the H{\footnotesize II} region N206 in the LMC, and so sometimes called ``the N206 SNR''. 
An elongated radio emission is observed between the SNR center and the east rim \citep{Klinger+02}.
A nonthermal X-ray emission is associated with this feature, suggesting the presence of 
a pulsar wind nebula and hence a CC origin of this SNR \citep{Williams+05}. 
Abundance measurements were made by Chandra and XMM-Newton observations \citep{Williams+05}. 
Our results are basically consistent with theirs.

B0532--67.5: 
The only preceding study of this SNR is performed by ROSAT \citep{Haberl+99}.
Although XMM-Newton has also observed this SNR, the result is still unreported. 
We therefore analyzed the XMM-Newton image as well, determining its angular size as given in table\,1. 
An X-ray spectrum with sufficient photon statistics has been obtained by Suzaku (i.e., this work), for the first time.

0103--72.6: 
This SNR is located with in the H{\footnotesize II} region DEM\,S125 (e.g., \cite{Filipovic+98}), 
which implies its massive star origin. The ASCA observation of this SNR confirmed its elevated 
metal abundances \citep{Yokogawa+02}, but the detailed chemical composition was not well constrained. 
The subsequent Chandra observation spatially resolved the central ejecta from the swept-up ISM shell \citep{Park+03}. 
The center spectrum indicates overabundances of O and Ne, consistent with the CC scenario. 
Our measurement also confirms the enhanced Ne abundance even in the integrated spectrum of the entire SNR.

0049--73.6: 
A spatially-resolved spectral analysis was performed in detail with Chandra \citep{Hendrick+05}.
The bright interior regions are dominated by O- and Ne-rich ejecta, suggesting a CC origin of this SNR,  
%No central source has, however, been detected. 
but the progenitor mass of $\gtrsim$ 40$\MO$ was ruled out from the lack of a large stellar-wind bubble 
(which is expected for a very massive star). Similarly to 0103--72.6, we confirm the high abundances of Ne 
from the spatially-integrated spectral data, whereas the Fe abundance is comparable to the mean SMC value.

0104--72.3 (a.k.a.\ IKT\,25): 
The progenitor type of this SNR has been controversial. 
Using XMM-Newton data, \citet{van-der-Heyden+04} derived an overabundance of Fe, and hence 
proposed an SN Ia origin. This was supported by a later Chandra observation by \citet{Lee+11}. 
However, using higher-statistics Chandra data, \citet{Lopez+14} found that the measured Ne/Fe 
abundance ratio is consistent with an aspherical CC SN of a massive progenitor. 
They therefore proposed that this SNR arose from a jet-driven bipolar SN explosion.
The star formation history at the periphery of this SNR also supports the CC scenario. 
Our analysis indicates a high Ne/Fe ratio, supporting the latest Chandra measurement by \citet{Lopez+14}. 
This will be discussed in more detail in the following subsections.

\subsection{Typing SNRs with the Fe/Ne Abundance Ratio}
\label{sec:Fe/Ne}

%%%--------------------------- 
\begin{figure}
  \begin{center}
        \includegraphics[width=8.0cm]{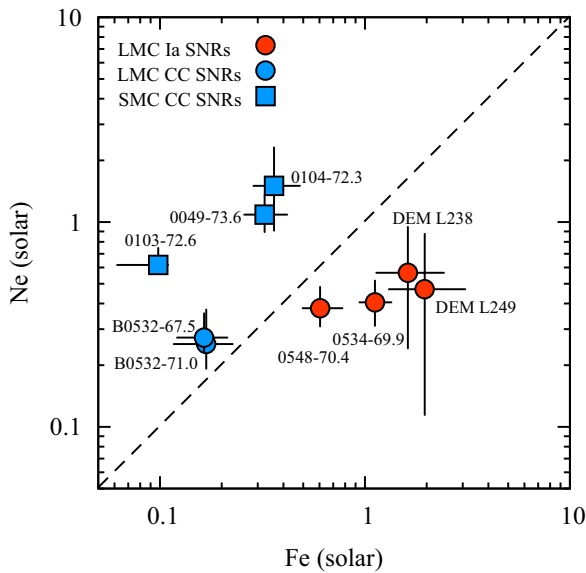}
  \end{center}
\caption{Measured abundances of Fe and Ne for the SNRs in the LMC (circles) and the SMC (squares). 
Red and blue represent candidates of Type Ia and CC SNRs, respectively.}
\label{fig:Fe-Ne}
\end{figure}

%This confirms that the SN ejecta still contribute to the X-ray spectra of these evolved remnants. 
As described above, we have confirmed the abundance enhancement of either Fe or the lighter elements 
(i.e., O, Ne and Mg) in most SNRs. For each, the progenitor type inferred from the abundance pattern agrees on the previous conclusions achieved by the spatially-resolved analysis with Chandra. This indicates that the integrated X-ray spectra 
of old (and thus ISM-dominant) SNRs are still contributed by the SN ejecta and thus can be used to 
discriminate the progenitor types with simple, unbiased spectral modeling. It should be noted that 
the ASCA observations of the middle-aged Magellanic SNRs (\cite{Hughes+98}; see also \S1) 
scarcely confirmed elevated metal abundances, although the soft X-ray luminosities of their targets 
($L_{0.5-5 {\rm keV}} \approx 10^{36-37}$\,ergs\,s$^{-1}$) were one or two order of magnitude higher than those 
of our samples (see table\,\ref{tab:fit}). The spectral sensitivity (i.e., effective area and energy resolution) is, 
therefore, essential for precise progenitor determination particularly for X-ray faint SNRs.

\begin{figure}
  \begin{center}
        \includegraphics[width=8.0cm]{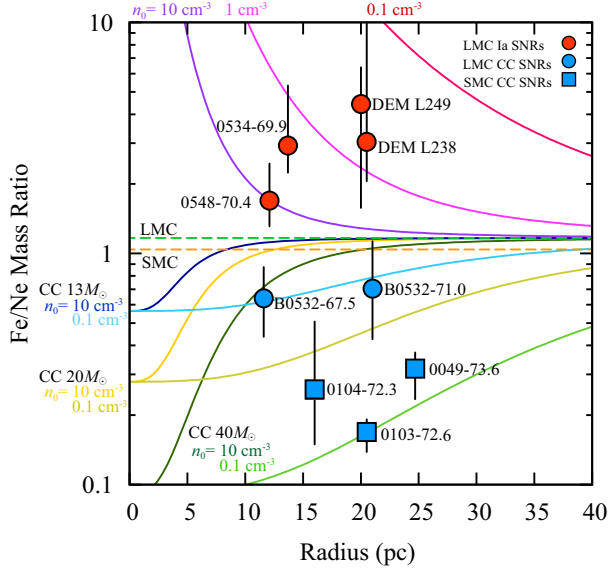}
  \end{center}
\caption{Fe/Ne mass ratios as a function of the SNR radius. 
The mean LMC and SMC mass ratios are indicated as the dashed lines. 
The solid curves show the theoretical mass ratios, where both ISM and ejecta contributions 
are taken into account (see text for more details). 
For the ISM component, uniform ambient densities ($n_0$) of 0.1--10\,cm$^{-3}$ are assumed. 
The metal masses of the SN Ia ejecta are assumed to be those of the WDD2 model \citep{Iwamoto+99}. 
For the CC SNRs, we use the values predicted by \citet{Kobayashi+06} for various progenitor masses 
between 13\,\MO and 40\,\MO with a metallicity of $Z$ = 0.004. 
}
\label{fig:Fe/NevsR}
\end{figure}

%We attempt to identify the best spectral feature to discriminate the progenitor type of the evolved SNRs 
The most prominent difference in the spectral feature among the nine SNRs (figure\,\ref{fig:spec}) is seen 
around 1\,keV, where either Fe L-shell or Ne K-shell emission is dominant. 
In figure\,\ref{fig:Fe-Ne}, we plot the measured Fe and Ne abundances (relative to  H) for each SNR, 
clearly separating the remnants into the two distinct groups. Given this fact, we propose 
the Fe/Ne mass ratio as the best quantity to distinguish the progenitor type of evolved SNRs 
(especially when a spatially-resolved spectrum is unavailable). 
The merit of this ratio is also justified from the theoretical point of view. 
First, Fe and Ne are the major products of Type Ia and CC SNe, respectively. 
Second, measurements of their abundance ratio are hardly affected by a foreground extinction or 
an electron temperature of plasmas, because of the similar transition energies of the Fe L-shell 
and Ne K-shell emission. 
Figure~\ref{fig:Fe/NevsR} shows the Fe/Ne mass ratio as a function of the SNR radius.  
All the values are significantly higher or lower than the mean LMC/SMC values for most SNRs.

%%%--------------------------- 

We here verify that the measured mass ratios make sense by comparing with theoretically-expected values 
in which the ISM contribution is taken into account. 
The total observed mass of Fe and Ne can be divided into those of the ISM and ejecta components as
\begin{equation}
	\frac{M_{\rm Fe}}{M_{\rm Ne}} = \frac{M_{\rm ISM, Fe}+M_{\rm ej., Fe}}{M_{\rm ISM, Ne}+M_{\rm ej., Ne}}.
	\label{eq.1}
\end{equation}
Assuming a uniform ambient density of $n_0$, 
the Fe (and similarly Ne) mass in the ISM component is calculated as 
\begin{equation}
	M_{\rm ISM, Fe} = m_{\rm Fe} \, n_0 \left(\frac{n_{\rm Fe}}{n_{\rm H}} \right)_{\rm LMC}  
						\left( \frac{4}{3} \pi R_{\rm SNR}^3 \right),  
\end{equation}
where $m_{\rm Fe}$ and $(n_{\rm Fe}/n_{\rm H})_{\rm LMC}$ are the mass of single Fe atom and 
the mean Fe/H number ratio of the LMC \citep{Russell+92}. 
For $n_0$, we consider various values in the range between 0.1\,cm$^{-3}$ and 10\,cm$^{-3}$. 
The assumption of the uniform ambient density should be valid for most Type Ia SNRs
(e.g., \cite{Badenes+07,Yamaguchi+14_FeKa}). A massive star like a red supergiant (RSG), 
on the other hand, usually explodes in a circumstellar matter (CSM) formed by its own 
pre-explosion stellar wind. 
However, given a typical wind velocity ($\sim$\,10\,km\,s$^{-1}$) and a length of the RSG stage 
($\sim$\,$10^5$\,yr), the CSM contribution should be significant only up to a few parsecs 
(e.g., \cite{Dwarkadas+05}), and thus the swept-up mass of the evolved SNRs should be 
dominated by the pre-existing ISM. The metal masses of the ejecta component are taken from 
the literature values of \citet{Iwamoto+99} and \citet{Kobayashi+06} for SNe Ia and CC SNe, respectively. 
The calculated mass ratios as a function of the SNR radius are shown as the solid curves in figure\,\ref{fig:Fe/NevsR}. 
We confirm that most SNRs (both Type Ia and CC) are in good agreement with the theoretical curves.

More specifically, the mass ratios observed in the Type Ia SNRs DEM\,L238 and DEM\,L249 suggest their low 
ambient density ($\lesssim 1$\,cm$^{-3}$), roughly consistent with the previous measurements \citep{Borkowski+06}. 
On the other hand, the SNR 0548--70.4 requires a relatively high density ($\sim 10$\,cm$^{-3}$), which is unusual 
for the LMC ISM. This implies that a part of the Fe ejecta is still unshocked or forms a low-density plasma. 
Several CC SNRs require a high progenitor mass. %0049--73.6
However, the absolute Fe ejecta mass of a CC SN is sensitive to a `mass cut' and thus known to be somewhat uncertain. 
In fact, we find that some other CC SN models (e.g., \cite{Thielemann+96}) predict a significantly lower Fe/Ne 
mass ratios for any progenitor mass. %for less massive progenitors (e.g., $\sim$25\MO). 
Moreover, our abundance measurements are based on the phenomenological analysis with simple spectral modeling. 
We caution, therefore, that the plot like figure~\ref{fig:Fe/NevsR} should not be used to determine the accurate 
progenitor mass of CC SNRs, although it is definitely useful for immediate SN type discrimination.

%%%---------------------------
\begin{figure}
  \begin{center}
        \includegraphics[width=8cm]{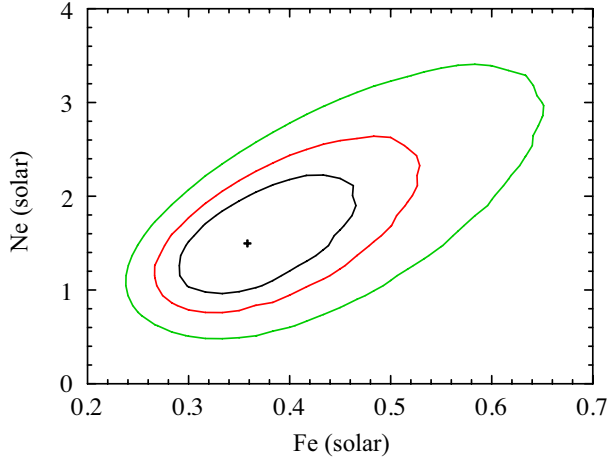}
  \end{center}
\caption{Confidence contour for the abundance ratio of Fe to Ne determined from the spectrum of 0104--72.3. 
Black, red, and green contours are given for confidence levels of 68\%, 90\%, and 99\%, respectively.}
\label{fig:0104--72.3_contour}
\end{figure}

The spectrum of 0104--72.3 is somewhat peculiar. 
Despite no obvious feature of Ne K emission, the derived Fe/Ne ratio is similar to the other CC SNRs. 
In figure\,\ref{fig:0104--72.3_contour}, we show confidence contours for the abundance ratios of 
Fe to Ne, which is almost consistent with the previous measurements by \citet{Lopez+14}. 
Unlike the other eight SNRs, 0104--72.3 does not require two-temperature components to fit the spectrum, 
but we confirm that an additional low-$kT_e$ component 
(with a fixed temperature between 0.1--0.3\,keV) does not change the Fe/Ne abundance ratio significantly. 
It would be worth noting that the fitting result of 0104--72.3 (figure\,1) shows no residuals around 
1.2\,keV that is commonly seen for the other Type Ia SNRs. 
This might be another piece of evidence for a low Fe abundance.

\subsection{Future Prospects}
\label{sec:prospect}

In this work, we have proved that an Fe/Ne mass ratio determined from a spatially-integrated soft X-ray spectrum of 
an evolved SNR clearly discriminates its progenitor type, when with adequate energy resolution of the detector. 
The X-ray mission ASTRO-H will be launched in early 2016, enabling the first high-resolution spectroscopy for 
largely-extended sources. The evolved Magellanic SNRs are suitable targets for this mission, because the typical 
spatial extent of the sources is too large to be observed with a grating spectrometer but is still smaller than 
the field of view of the Soft X-ray Spectrometer (SXS) aboard ASTRO-H ($3' \times 3'$: \cite{Takahashi+14}).

%%%--------------------------- 
\begin{figure*}
  \begin{center}
        \includegraphics[width=16cm]{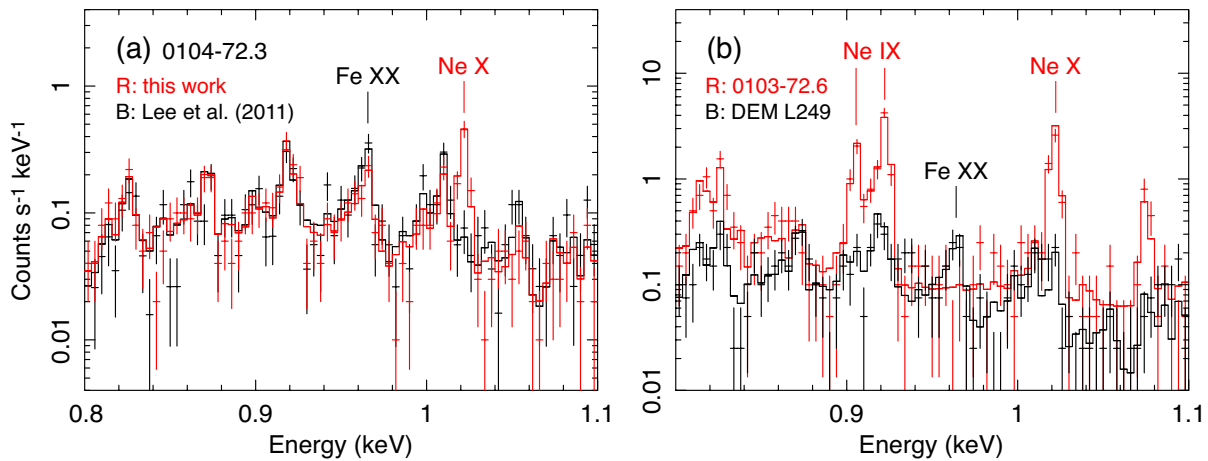}
  \end{center}
\caption{(a) Simulated spectra of SNR 0104--72.3 in the 0.8--1.1\,keV band for observations with 
the SXS aboard ASTRO-H. We use parameters of our best-fit model (see table \ref{tab:fit}) and the Fe/Ne 
abundance ratio reported by \citet{Lee+11} for the red and black data points, respectively. 
The assumed exposure time is only 25\,ks for both cases. We can easily distinguish the two cases 
with this short exposure. \ 
(b) Simulated SXS spectra of DEM\,L249 (black) and 0103--72.6 (red) with the assumed exposures 
of 10\,ks and 5\,ks, respectively. 
}
\label{fig:sxs}
\end{figure*}

Figure\,\ref{fig:sxs}a shows simulated SXS spectra of the SNR 0104--72.3 with an exposure of only 25\,ks,  
where we assume our best-fit model (black) or the Fe/Ne abundance ratio reported by \citet{Lee+11} (red). 
The two cases (i.e., CC and Type Ia progenitor scenarios) can easily be distinguished with this short exposure time, 
since the strong Ly$\alpha$ emission expected only for the CC scenario. 
For the other `normal' SNRs with the highest or lowest Fe/Ne ratios among our samples (i.e., DEM\,L249 and 0103--72.6),
 we can resolve the key spectral features with even shorter exposure time (5--10\,ks) 
as demonstrated in figure\,\ref{fig:sxs}b. A systematic study of evolved SNRs is, therefore, promising for ASTRO-H.
With an ultimate X-ray mission, like Athena\footnote{The Athena mission is supposed to have a $\sim$100 times larger 
effective area than ASTRO-H, compensating for flux decrease due to the large distance to the outer galaxies.
Furthermore, because of the large distance (plus the high angular resolution of the X-ray mirror), multiple SNRs 
can be detected and resolved with a single pointing observations.}, 
we will moreover be able to perform similar studies for SNRs in other nearby galaxies (e.g., M31). 
This will help understand the detailed chemical evolution in an entire spiral galaxy.

Finally, we should mention that the Fe/Ne-ratio diagnostic presented in this work might be valid only for 
Sedov-phase SNRs where the entire ejecta have already been shock-heated. In young Type Ia SNRs, 
on the other hand, a significant fraction of the Fe ejecta is still cool and/or in very low ionization states. 
In fact, Fe L emission from SN\,1006 is almost invisible for this reason (\cite{Vink+03,Yamaguchi+08}). 
For young SNRs, diagnostics using other elements (e.g., \cite{Hughes+95}), X-ray morphology (\cite{Lopez+11}), 
or ionization states of the ejecta (\cite{Yamaguchi+14_FeKa}) would be more useful and robust.

%+++++++++++++++++++++++++++++++++++
\section{Conclusions}
\label{sec:conclusion}

We have presented systematic analysis of nine SNRs in the Magellanic Clouds 
(DEM\,L238, DEM\,L249, 0534--69.9, 0548--70.4, B0532--71.0, B0532--67.5, 
0103--72.6, 0049--73.6, and 0104--72.3) observed with the Suzaku/XIS. 
These are all evolved SNRs with an age of $\gtrsim$\,10,000\,yr, so their integrated 
X-ray spectra in the soft X-ray band are dominated by the swept-up ISM. 
Nevertheless, we confirm clear signatures of the ejecta from all the observed targets. 
%The prominent difference among their spectra is seen around 1\,keV, where either 
%Fe L-shell or Ne K-shell emission is dominant. 
The Fe/Ne mass ratio derived from our simple unbiased spectral analysis clearly 
discriminates the progenitor types. DEM\,L238, DEM\,L249, 0534--69.9, and 0548--70.4
are classified as Type Ia remnants, while the other five are CC SNRs, all consistent 
with the previous Chandra results based on the spatially-resolved analysis. 
Since Fe L-shell and Ne K-shell lines are fall in the same energy range, measurements 
of their mass ratio are hardly affected by a foreground extinction or plasma temperature---this is 
the remarkable advantage of this simple diagnostic.  
Future high-resolution spectroscopy with the ASTRO-H/SXS will resolve 
the key spectral features much more clearly with very short exposure, 
enabling robust systematic studies of the Magellanic Sedov-phase SNRs. 
This helps reveal detailed history of the recent chemical evolution in an outer galaxy.

\begin{ack}
We thank all the members of the Suzaku operation team for their continuous effort to develop and maintain the satellite. 
Y.T. was supported by Grant-in-Aid for the Promotion of Science (JSPS) Fellows (No. 25$\cdot$5448).
\end{ack}

\end{document}